\documentclass[aps,a4paper,twocolumn]{revtex4-2}
\usepackage{amsmath}
\usepackage{graphicx}
\usepackage{subfigure}
\usepackage[usenames,dvipsnames]{color}
\definecolor{darkblue}{RGB}{0,0,196}
\usepackage[colorlinks=true,linkcolor=darkblue,citecolor=darkblue,urlcolor=darkblue]{hyperref}
\usepackage{setspace}
\usepackage{url}
\usepackage{hyperref}
\usepackage{ulem}
\usepackage{xcolor}
\hypersetup{
  colorlinks   = true, %Colours links instead of ugly boxes
  urlcolor     = red, %Colour for external hyperlinks
  linkcolor    = blue, %Colour of internal links
  citecolor   = blue %Colour of citations
}
\usepackage{graphicx}
\usepackage{amsmath,bbm}
\usepackage{amssymb,bm}
\usepackage{yfonts}
\usepackage{comment}

\usepackage[usenames,dvipsnames]{color}
\definecolor{darkblue}{RGB}{0,0,196}
\usepackage[colorlinks=true,linkcolor=darkblue,citecolor=darkblue,urlcolor=darkblue]{hyperref}
\begin{document}
\title{Emergent spin polarization from $\rho$ meson condensation in rotating hadronic matter}

\author{Kshitish Kumar Pradhan$^{1}$} 
\author{Dushmanta Sahu$^{2}$} 
\author{Raghunath Sahoo$^{1}$} 
\email{Corresponding Author: Raghunath.Sahoo@cern.ch}
\affiliation{$^{1}$Department of Physics, Indian Institute of Technology Indore, Simrol, Indore 453552, India}
\affiliation{$^{2}$ Instituto de Ciencias Nucleares, Universidad Nacional Autónoma de México, Apartado Postal 70-543, México Distrito Federal 04510, México}

\begin{abstract}
The behavior of vector mesons in extreme environments provides a unique probe of non-perturbative Quantum Chromodynamics. We investigate the conditions for Bose-Einstein condensation (BEC) of spin-1 $\rho$ mesons in dense rotating hadronic matter, a regime relevant to the peripheral heavy-ion collisions and the interiors of rapidly rotating neutron stars. When the $\rho$ meson chemical potential ($\mu_\rho$) approaches its effective mass ($m_\rho^*$), a phase transition to BEC occurs. We demonstrate that this transition is non-trivially influenced by global rotation, which couples to the spin of the $\rho$ mesons, leading to a macroscopic spin alignment of the condensate along the axis of rotation. This interplay between condensation and rotation results in distinct polarization patterns, which can serve as a possible signature of a BEC in experiments. The results suggest that rapidly rotating neutron stars may harbor an anisotropic, spin-polarized $\rho$-condensed phase, which could impact their equation of state. % and gravitational wave signatures.

\end{abstract}
\maketitle
\section{Introduction}

The study of strongly interacting matter under extreme conditions of temperature, density, and rotation is a central pursuit in high-energy nuclear physics. Non-central heavy-ion collisions in laboratories such as the Relativistic Heavy Ion Collider (RHIC) and the Large Hadron Collider (LHC) create a novel state of matter called Quark-Gluon Plasma (QGP), endowed with immense angular momentum, leading to the strongest vortical fields known in nature~\cite{STAR:2017ckg}. Simultaneously, the dense cores of neutron stars provide a natural astrophysical setting where high baryon density and rapid rotation combine, influencing the equation of state and observable phenomena like pulsar glitches and gravitational waves~\cite{Haensel:2007yy, Paschalidis:2016vmz}. These diverse systems offer complementary avenues to explore the rich phase structure of quantum chromodynamics.

Among the various probes of this exotic matter, vector mesons, particularly the $\rho$ meson ($J^{PC} = 1^{--}$) serve as important probes. Due to their short lifetime and strong coupling to the dense medium, their in-medium spectral properties are a sensitive indicator of chiral symmetry restoration and collective effects~\cite{Rapp:2009yu, Hayano:2008vn}. While the condensation of pseudoscalar pions ($\pi$, $J^{PC} = 0^{-+}$) has been extensively studied~\cite{Begun:2006gj, Begun:2015ifa, Begun:2008hq, Brandt:2017oyy, Kuznietsov:2021lax}, the potential Bose-Einstein Condensation (BEC) of their vector counterparts, the $\rho$ mesons, remains a significantly less explored frontier. This possibility arises in high-baryon-density environments where the $\rho$ meson chemical potential $\mu_\rho$ is driven to match its effective mass $m_\rho^*$, a condition which can be attained in low-energy heavy-ion collisions or inside the neutron star~\cite{Voskresensky:1997ub, Mallick:2014faa}. In such regimes, a strong magnetic field can stabilize the $\rho$ condensate against the decays of $\rho$ mesons~\cite{Chernodub:2010qx}. The $\rho$ condensate can couple to the background magnetic field, forming a spin-aligned, superconducting state as proposed for QCD in strong magnetic fields in Ref.~\cite{Chernodub:2010qx, Chernodub:2011mc}.

Rotation adds an interesting and new dimension to this picture. The vorticity generated in off-central heavy-ion collisions can reach $\omega \sim 10^{22}$ s$^{-1}$, corresponding to $\hbar \omega \sim 10^{-2}$ GeV~\cite{STAR:2017ckg}. Unlike the well-studied pion condensates, $\rho$ mesons carry spin-1, introducing a rich polarization structure that couples to the medium's angular momentum and electromagnetic fields. In rotating systems, such as the vortical quark-gluon plasma produced in off-central heavy-ion collisions or the superfluid vortices in neutron stars, this coupling can lead to a spin alignment of $\rho$ mesons along the rotation axis. The resulting polarization patterns may be further modified by the presence of a $\rho$ BEC, where the macroscopic occupation of the ground state amplifies collective spin effects. These dynamics are not only of intrinsic interest for QCD but also bear direct significance on astrophysical observables. The presence of a $\rho$ meson condensate could soften the neutron star equation of state, while polarized $\rho$ mesons might contribute to anisotropic neutrino emission or leave imprints in gravitational wave signals from merging compact objects.

In literature, pion BEC has been studied extensively. Refs.~\cite{Begun:2008hq} show the properties of a pion BEC occurring in the thermodynamic and finite volume limit. In Ref.~\cite{Savchuk:2020yxc}, the authors explore the effect of repulsive interaction on the pion BEC. Moreover, the effect of condensation on the transport properties of the pion has also been studied~\cite{Pradhan:2022mig}. In Ref.~\cite{Deb:2021gzg}, the authors explore the effect of non-extensivity in pion BEC. However, little has been explored in the area of rho-meson BEC. In Ref.~\cite{Brauner:2016lkh}, the authors study the rho meson condensation using the Nambu-Jona Laisino model and the linear sigma model. Similarly, in Refs.~\cite {Shivam:2019cmw, Mallick:2014faa, Voskresensky:1997ub}, the authors investigate the possibility of rho meson condensation in neutron stars. In addition, a new avenue has recently opened up regarding the possibility of pion condensation in rotating matter under an external magnetic field, where the authors show that rotation and a magnetic field help in faster condensation~\cite{Liu:2017spl}. However, no study has been done to explore the effect of rotation and condensation on the polarization pattern of the particles under consideration.

In this article, we bridge this gap by presenting a systematic study of the interplay between BEC and spin polarization for $\rho$ mesons in a dense, rotating hadronic gas. We establish the phase boundary for $\rho$ meson condensation and demonstrate how global rotation not only facilitates the condensation but also imprints a definitive polarization signature on the condensed phase. Our findings have direct implications; in heavy-ion collisions, the predicted polarization anisotropy should be observable through angular distributions of dilepton decays, providing a clean signal of this novel phase. In astrophysics, a spin-polarized $\rho$ meson condensate could lead to a softer equation of state for neutron stars and contribute to anisotropic pressure, leaving imprints in the pre- and post-merger gravitational wave signals of binary compact star collisions~\cite{Kolomeitsev:2017gli, Sedrakian:2022ata, Bauswein:2018bma}. This work thus establishes a new link between the dynamics of relativistic heavy-ion collisions and the microphysics of dense astrophysical compact objects.

\section{Formalism}

We start with the single-particle Bose-Einstein (BE) distribution function, which is given by~\cite{KHuang},
\begin{equation}
%f = \frac {1}{{\rm exp}\left(\frac{E - \mu}{T}\right) - 1}.
f = \left[{\rm exp}\left(\frac{E - \mu}{T}\right) - 1\right]^{-1},
\label{eq_distfun}
\end{equation}
where $E$ is the energy of the particle given by $E = \sqrt{k^2+m^2}$, $k$ and $m$ being the momentum and mass of the particle. Here, $\mu$ is the chemical potential of the particle, whereas $T$ is taken as the temperature of the system. Considering a gas of rho mesons having mass $m$, the total number density ($n_{\rm tot}$) can be written as \cite{Pradhan:2022mig}
\begin{equation}
\label{thlim}
\begin{split}
    n_{\rm tot} &= \frac{g}{(2\pi)^3} \int \frac{d^{3}k}{{\rm exp} \left(\frac{\sqrt{k^2 + m^2} - \mu}{T}\right) - 1} \\
    &\approx \frac{gTm^2}{2\pi^2}\sum_{j=1}^{\infty}\frac{1}{j}K_2(jm/T)\exp(j\mu/T),
\end{split}
\end{equation}
where $g$ is the spin degeneracy of the rho meson, $K_2$ is the modified Bessel function and $j$ being a positive integer. In normal conditions, this is always valid for $\mu<m$. One interesting feature of the BE distribution function is that, as $\mu\to m$, the ground state starts crowding and diverges at $\mu=m$. We can split the total density as, 
\begin{align}
    \label{eq_totln}
    n_{\rm tot} &= n_{\rm cond} + n_{\rm ex}\\
    =&\frac {g}{ V[{\rm exp} (\frac{m - \mu}{T}) - 1]} + \int \frac{d^3k}{(2\pi)^3} \frac {g}{{\rm exp} \bigg(\frac{\sqrt{k^2 + m^2} - \mu}{T}\bigg) - 1}.
\end{align}
Here, the first term is the occupation density of the ground state, whereas the second term, known as the thermal component, constitutes all particles in the excited state. The BEC transition happens for $\mu=m$ (at temperature $T=T_c$), when the maximum number of particles in the excited states is reached and is equal to the total density of the system. Therefore, considering a fixed density, the BEC transition temperature ($T_c$) is defined by \cite{Pradhan:2022mig, chapter3Y, megias2022, chapter12SM}
\begin{equation}
    \label{eq_Tc}
    n_{\rm ex}(T=T_c, \mu = m) = n_{\rm tot}.
\end{equation}

In the presence of a rotation ($\omega$) in the system, say along the $z$ direction, the energy dispersion relation gets shifted by a factor proportional to the rotation as $\varepsilon_{l} = E-(l+s)\omega$, where $s$ is the $z$ component of spin, and $l$ is the orbital angular momentum quantum number along the $z$-axis. The pressure for a Bose gas in the presence of rotation for $ith$ species can therefore be written as~\cite{Fujimoto:2021xix, Mukherjee:2023qvq}
\begin{align}
\label{eq_pressure}
    P_i &= -\frac{T}{8\pi^2} \sum_{\ell=-\infty}^\infty \int dk_r^2 \int  dk_z\; \sum_{\nu = \ell}^{\ell + 2S_i}  J_\nu^2(k_r r) \notag\\
    &\qquad\qquad \times \log\left\{1-\exp[-(\varepsilon_{\ell,i} - \mu_i) / T]\right\},
\end{align}
where the energy of each particle in the presence of rotation is given by, $\varepsilon_{l,i}$ = $\sqrt{k_r^2 +k_z^2 +m_i^2}-(l+s)\omega$. The $r$ is the radial distance from the axis of the rotation, whereas $J_\nu$ are the Bessel functions of the first kind. The radial and longitudinal momenta are given by $k_r$ and $k_z$, respectively. The causality condition requires imposing a boundary condition at $r=R$ such that $R\omega\leq1$ ($c=1$ in natural units). We have considered $R$ = 5 fm throughout this study. This boundary condition led to the quantization of the angular momentum as $k_r=\xi_{l,i}/R$~\cite{Fujimoto:2021xix}, with $\xi_{l,i}$ being the $ith$ zero of the Bessel function such that, $J_l(\xi_{l,i})=0$. The discretization effect of momenta affects especially the lower momenta region, and therefore, the $k_r$ integration faces a lower bound = $\xi_{l,i}/R=\xi_{l,i}\omega$. It is important to emphasize that the formalism has a radial dependence through the Bessel function, $J_\nu(k_r r)$. However, to extract and interpret the genuine rotational effects and to avoid any ambiguity because of spatial variations across the radius, we remove this radial dependence by evaluating all thermodynamic quantities at a fixed value of $r$ as previously done in~\cite{Fujimoto:2021xix, Mukherjee:2023qvq, Mukherjee:2023ijv}. This removes any additional complications arising from spatial inhomogeneity, which can be explored in detail elsewhere. Given that the centrifugation effects will be maximum at the edge of the system, all the results here are obtained by fixing $r=R$~\cite{Mukherjee:2023ijv}.

Considering a gas of $\rho$ mesons, one can obtain total number density from Eq.~(\ref{eq_pressure}) as 
\begin{align}
    \label{eq_numden}
    n_{\rm tot} &= \bigg(\frac{\partial P}{\partial \mu}\bigg)_{T,\omega}\nonumber \\
    &= \frac{1}{8\pi^2} \sum_{\ell=-\infty}^\infty \int dk_r^2 \int  dk_z\; \sum_{\nu = \ell}^{\ell + 2S_i}  J_\nu^2(k_r r) \notag\\
    &\qquad\qquad \times \frac{1}{\exp[(\varepsilon_{\ell,i} - \mu_i) / T]-1}.
\end{align}

Note that the summation over $\nu$ is considered in Ref.~\cite{Fujimoto:2021xix} to take care of the spin degeneracy for simplicity. However, in the presence of rotation, the degeneracy lifts up, and depending on the coupling of different spins with the direction of angular momentum, the density varies with spin. In that case, for a study involving the spin characteristic of the particles, the $\nu$ summation must be ignored. Instead, there must be a summation over each spin component of the considered particle. Here, for the rho meson, the summation goes over $s=-1$ to $s=+1$.

In a non-rotating rho meson gas, Bose–Einstein condensation (BEC) occurs when the chemical potential equals the ground-state energy, i.e., $\mu = \varepsilon_0 = m_\rho$~\cite{Pradhan:2022mig}. For charged $\rho^\pm$ mesons in dense matter, the chemical potential arises from the isospin chemical potential as $\mu_\rho = \pm\mu_I$. In neutron-rich environments, as in neutron stars, the in-medium $\rho$ mass decreases with increasing density and magnetic field, leading to condensation at a chemical potential, a few times the nuclear saturation density~\cite{chapter12SM, Shivam:2019cmw, Mallick:2014faa}. In the present case of a rotating system with constant $\rho$ mass, we treat $\mu$ as an effective chemical potential incorporating both iso-spin and rotational contributions. Rotation modifies the single-particle energies, effectively shifting them, while the cylindrical boundary at $r = R$ quantizes the radial momentum. Consequently, the ground-state energy $\varepsilon_0$ differs from $m_\rho$ as for the ground state $k_r \neq 0$, but is finite and determined by the boundary condition. Therefore, BEC in a rotating $\rho$ gas occurs when $\mu = \mu_c = \varepsilon_{\text{min}}$, where $\varepsilon_{\text{min}}$ denotes the minimum single-particle energy. The derivation of this minimum energy $\varepsilon_{\text{min}}$ is central to establishing the BEC condition in a rotating, confined system. From Eq.~(\ref{eq_numden}), we have the expression for the particle's energy as
\begin{equation}
    \label{eq_energyi}
    \varepsilon_{l} = \sqrt{k_r^2 +k_z^2 +m^2}-(l+s)\omega,
\end{equation}
where $s$ is the $z$ component of the particle's spin. To find the absolute ground state, we minimize this energy with respect to all degrees of freedom. First, the longitudinal momentum is set to zero, $k_z = 0$, to eliminate kinetic energy along the axis. Second, for a $\rho$ meson with spin projections $s = -1, 0, +1$, the coupling $-(l+s)\omega$ is minimized by selecting the maximum value $s=+1$ to gain the most energy from rotation. For a rotating rho meson gas, the minimum or ground state energy (with $k_z$ = 0, and $\varepsilon_{l}-\mu\geq$ 0 in mind) is then given by
\begin{equation}
    \label{eq_energy0}
    \varepsilon_{min} = \sqrt{(\xi_{l_{max},1}\omega)^2+m^2} - (l_{max}+1)\omega.
\end{equation}
where $l_{\text{max}}$ is the value that minimizes the expression. The BEC transition occurs when $\mu = \varepsilon_{\text{min}}$. Therefore, the condition of the critical temperature for the onset of BEC of rotating $\rho$ meson gas is obtained by 
\begin{equation}
    \label{eq_Tc_omega}
    n_{\rm ex}(T=T_c, \mu = \varepsilon_{min}) = n_{\rm tot}.
\end{equation}
Below $T_c$, the condensate fraction can be estimated as~\cite{Begun:2008hq, Pradhan:2022mig}
\begin{align}
    \label{eq_nc_nt}
    &n_{\rm tot} = n_{\rm cond} + n_{\rm ex}, \nonumber \\
    \Rightarrow &\frac{n_{\rm cond}}{n_{\rm tot}} = 1-\frac{n_{\rm ex}}{n_{\rm tot}} = 1-\frac{n_{ex}(T<T_c, \mu=\varepsilon_{min})}{n_{ex}(T=T_c, \mu=\varepsilon_{min})}.
\end{align}

In a non-rotating gas, the density is spin degenerate. However, for a rotating gas of $\rho$ mesons, there is an imbalance between the states owing to their nature of coupling along the direction of angular momentum. To study the polarization effect, we examine the spin density of each state. In a more general quantum-statistical description, the spin state of a vector meson ($S=1$) is represented by the spin density matrix $\rho_{mm'} = \langle S,m_s|\hat{\rho}|S,m_s'\rangle$, whose diagonal elements $\rho_{11}$, $\rho_{00}$, and $\rho_{-1-1}$ denote the occupation probabilities of the spin substates with projections $m_s=+1, 0, -1$, respectively~\cite{Liang:2004xn, Becattini:2013fla}. One commonly measured spin alignment parameter is $\rho_{00}$ $\sim n_0 / (n_{+1} + n_0 + n_{-1})$. This $\rho_{00}$ quantifies the probability of the meson spin being aligned perpendicular to the quantization axis. A measured value of $\rho_{00}\neq1/3$ signals a breakdown of statistical spin population and is a direct experimental signature of global polarization~\cite{ALICE:2019aid, STAR:2022fan}. %and is used as an experimental observable to determine the polarization of produced particles~\cite{}. 
Theoretically, in the quark level, it is related to quark polarization through~\cite{Liang:2004xn, Sun:2024anu}
\begin{align}
    \rho_{00} = \frac{1 - P_q P_{\bar q}}{3 - P_q P_{\bar q}},
\end{align}
where the quark polarization $P_q$ is given as
\begin{align}
    P_{q} = \frac{n_{q_\uparrow}-n_{q_\downarrow}}{n_{q_\uparrow}+n_{q_\downarrow}},
\end{align}
where $n_{q_\uparrow}$ and $n_{q_\downarrow}$ are the quark densities of spin up and spin down states, respectively. Therefore, by analogy, we can study the polarization at the hadronic level by studying their spin alignment. A deviation of $\rho_{00}$ from the statistical limit of $1/3$ indicates the presence of spin alignment induced by vorticity or other medium effects, while a nonzero $P$ measures the vector polarization of the gas. 
Our formalism, based on relativistic quantum statistics with rotation, provides a robust and well-established first insight into the expected phenomena, setting the stage for future, more complex effective field theory studies.
\begin{figure*}[ht!]
\begin{center}
\includegraphics[scale = 0.44]{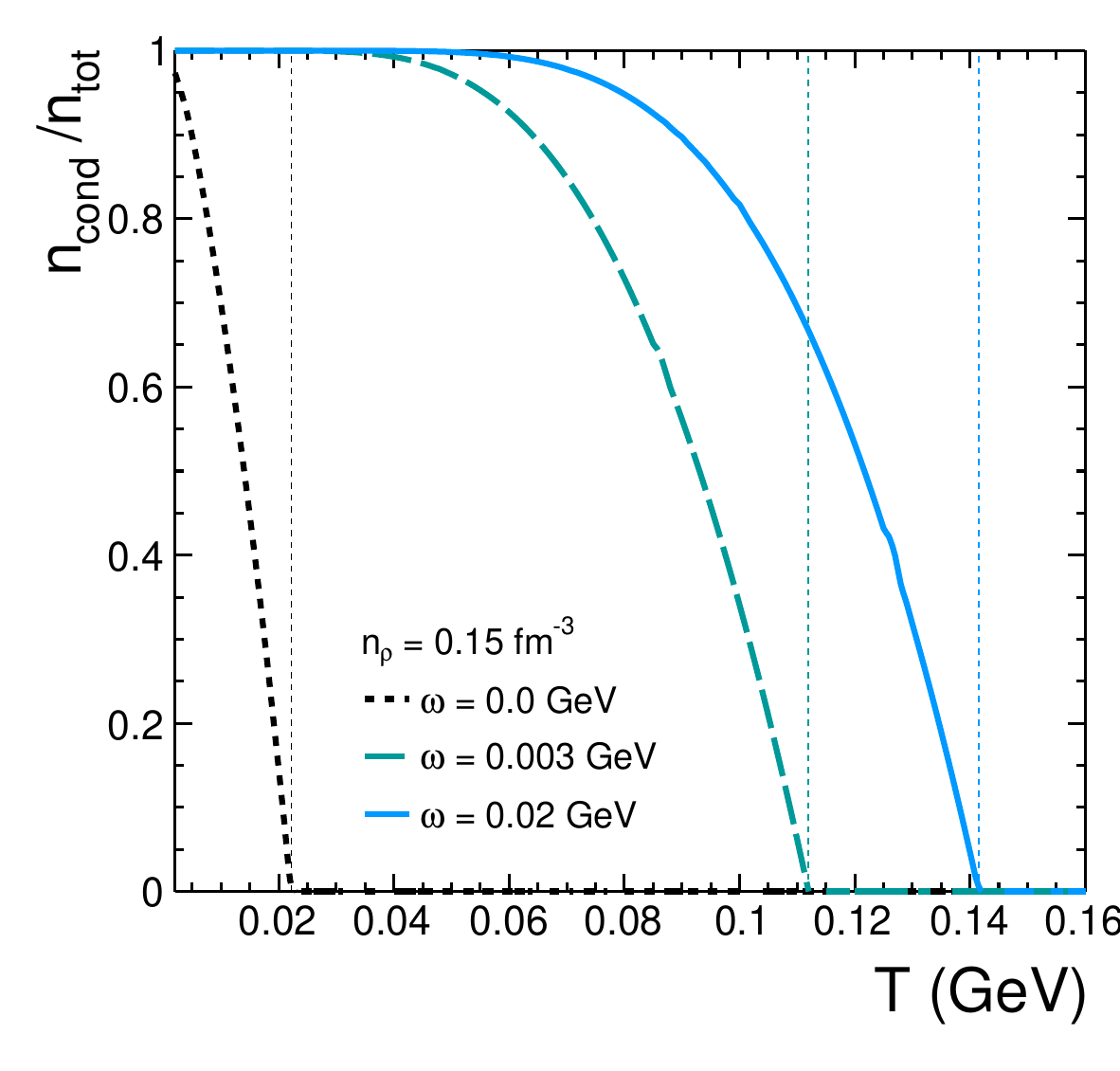}
\includegraphics[scale = 0.44]{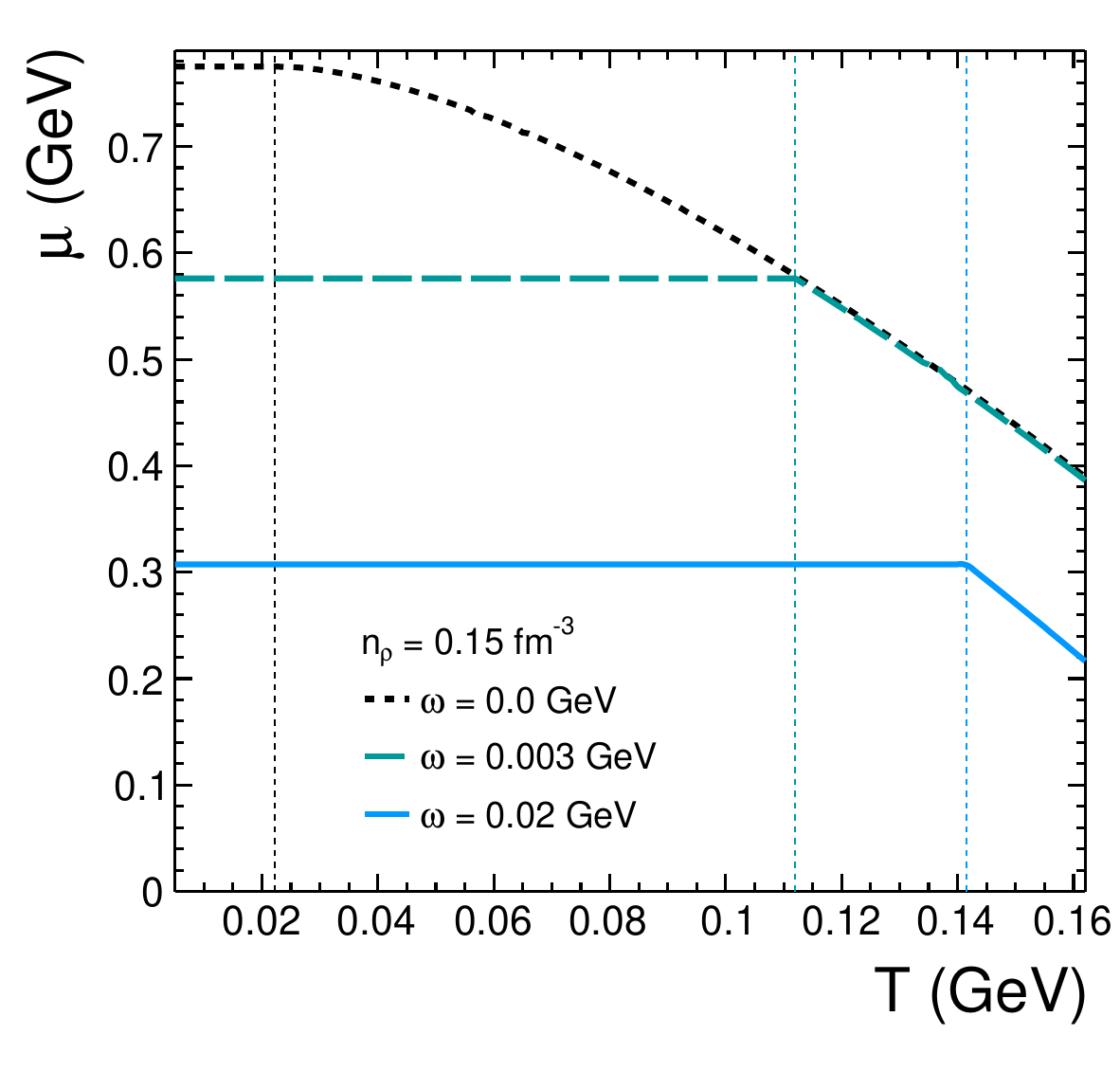}
\caption{(Colour Online) The left panel shows the condensate fraction as a function of temperature, and the right panel demonstrates the temperature dependence of chemical potential for the three distinct values of $\omega$.}
\label{fig1}
\end{center}
\end{figure*}

\section{Results and Discussion}

In this work, we have considered a total density of rho meson gas as $n_{\rm tot}$ = 0.15 fm$^{-3}$. This accounts for all the spin states of the $\rho$ meson, namely the spin-up, spin-down, and spin-zero components within a cylindrical rotating system of fixed radius $R$ = 5 fm. This value of $R$ is motivated by various experimental and theoretical estimates of characteristic system sizes, which are approximately of the order 4 to 9 fm~\cite{ALICE:2013uhj, ALICE:2020mkb, Sahu:2019tch}. In the absence of rotation ($\omega = 0$), the system is symmetric, and all three spin states possess equal densities, each contributing $0.05~\text{fm}^{-3}$. At finite rotation, this degeneracy is lifted, leading to a redistribution of density among the spin states. With an increase in rotation, the up-spin density increases, while the spin-down states decrease, keeping spin-zero states almost unchanged, such that the total density is fixed at 0.15 fm$^{-3}$. Let us first look at the effect of rotation on the BEC of the $\rho$ meson gas. Eq.~(\ref{eq_nc_nt}) implies that below $T_c$, as $T\to 0$, the condensate fraction approaches unity. This means the total density considered here ($n_{\rm tot}$ = 0.15 fm$^{-3}$) is in the condensate state. With an increase in $T$, the condensate fraction decreases, and at $T=T_c$, the condensate vanishes, leaving behind all the particles to be in the excited state. In the left panel of Fig.~\ref{fig1}, we show the condensate fraction as a function of temperature for three different values of $\omega$, including the non-rotating case, for which $\omega$ = 0. As expected, for all cases, the condensate fraction is zero at $T_c$ and increases as one goes towards low $T$, approaching unity as $T$ tends to zero. With an increase in the magnitude of rotation, the BEC transition temperature, $T_c$, is observed to increase, which can be understood by studying the interplay of $T$, $\omega$, and $\mu$ with the constraint of fixed number density. The right panel of Fig.~\ref{fig1} shows the temperature dependence of the chemical potential of the system. At zero rotation, the system reduces to the simple case of a non-rotating relativistic Bose gas for which the BEC happens when $\mu=m$. This behavior is confirmed in the right panel of Fig.~\ref{fig1}. At zero rotation, $\mu=m=0.775$ GeV for $T<T_c$, and at high temperature (for $T>T_c$), the requirement of fixed number density results in a decrease in $\mu$, consistent with the expected thermodynamic behavior~\cite{Pradhan:2022mig, Begun:2008hq}.

At finite rotation, following Eq.~(\ref{eq_energy0}), it is evident that the effective chemical potential of the system is reduced, with the reduction being proportional to the angular velocity. This behavior is particularly interesting, as it implies that rotation itself facilitates the onset of BEC by lowering the chemical potential required for condensation compared to the non-rotating case. Consequently, this also explains the rise in critical temperature $T_c$ with rotation, given the constraint of a fixed $n_{\rm tot}$. 
\begin{figure*}[ht!]
\begin{center}
\includegraphics[scale = 0.44]{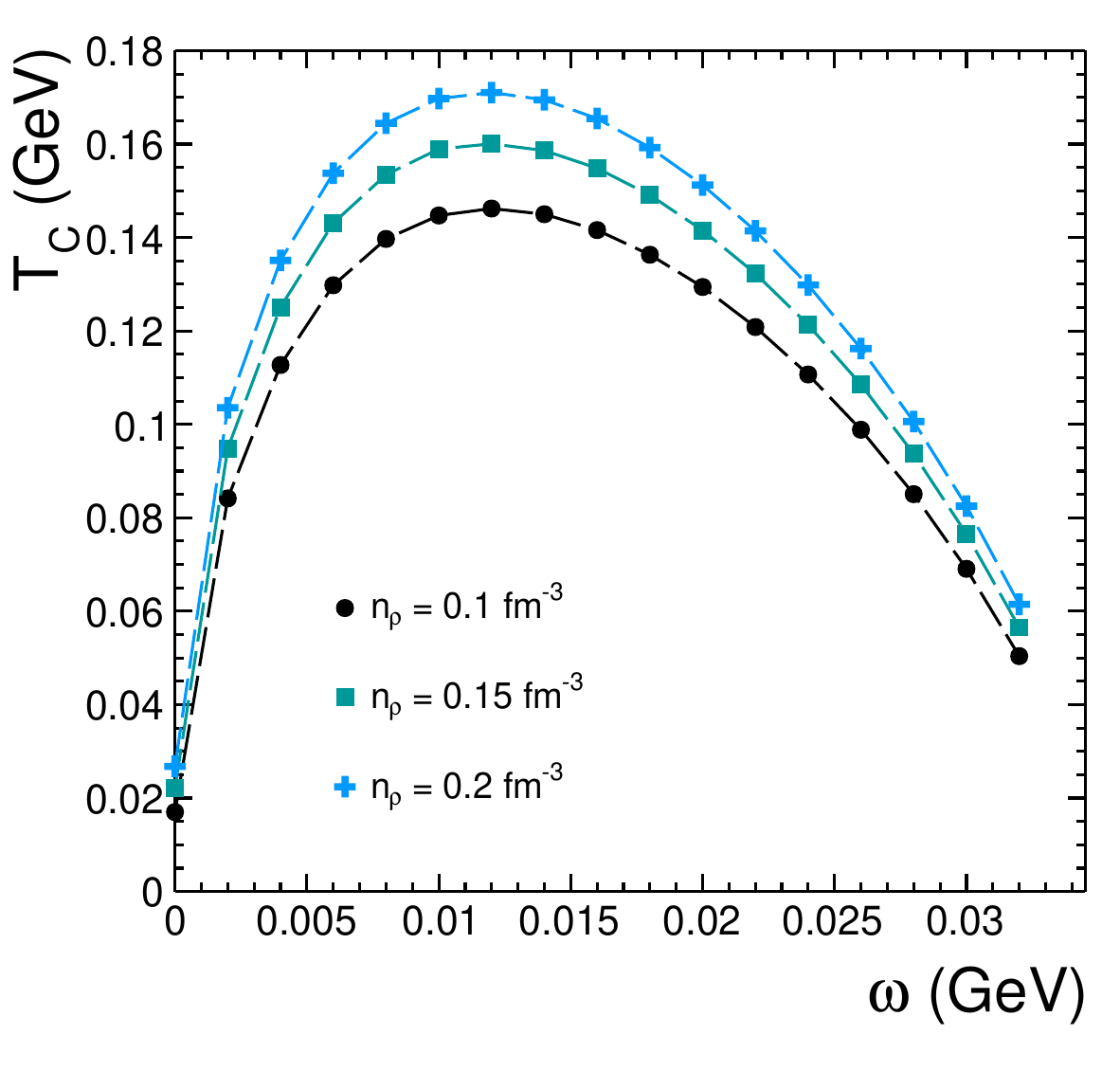}
\includegraphics[scale = 0.44]{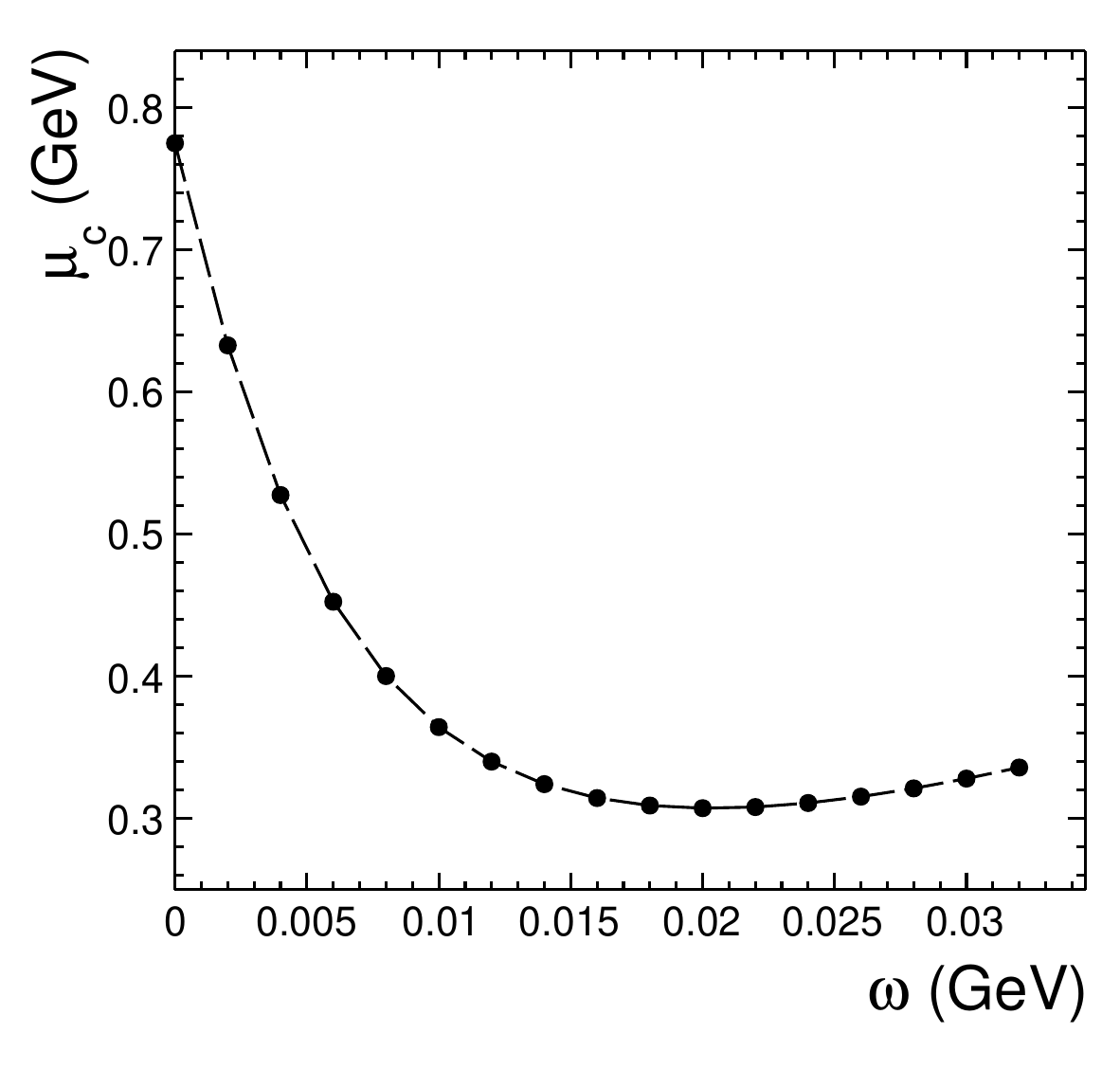}
\caption{(Colour Online) The variation of BEC transition temperature (left panel) and transition chemical potential (right panel) with rotation ($\omega$) for a constant rho meson density.}
\label{fig2}
\end{center}
\end{figure*}
For different total densities considered, the variation of $T_c$ with $\omega$ is shown in Fig.~\ref{fig2}. It is evident from the figure that the critical temperature increases with an increase in number density (at a particular $\omega$), which is in line with previous studies, such as in~\cite{Begun:2008hq}. However, with an increase in $\omega$, the $T_c$ increases rapidly and then decreases after attaining a maximum. The maxima of $T_c$ observed are found to be independent of the total density considered. We can understand this non-monotonic behaviour of $T_c$ with $\omega$ by analyzing the minimum energy $\varepsilon_{\text{min}}$ required to achieve the BEC condition in a rotating system. The minimization over the orbital quantum number $l$ in Eq.~(\ref{eq_energy0}) reveals a competition that governs $T_c(\omega)$: while the term $-(l+1)\omega$ favors large $l$, the centrifugal term $\sqrt{(\xi_{l, 1} \omega)^2 + m^2}$ grows with $l$ as $\xi_{l, 1}$ increases. The non-monotonic $T_c(\omega)$ emerges from this expression, at moderate $\omega$, the linear $-(l+1)\omega$ term dominates, significantly lowering $\varepsilon_{\text{min}}$ and thereby raising $T_c$. However, at high $\omega$, the quadratic centrifugal term $(\xi_{l, 1} \omega)^2$ becomes dominant, increasing $\varepsilon_{\text{min}}$ and thus suppressing $T_c$. %The maximum $T_c$ occurs at the critical $\omega$ where the slope of $\varepsilon_{\text{min}}$ changes sign. 
We obtain $\mu_c$ (=$\varepsilon_{min}$) as a function of $\omega$ for different $n_{\rm tot}$. As expected, the chemical potential required to achieve BEC ($\mu_c$) does not depend on $n_{\rm tot}$. For example, at $\omega=0$ GeV, the $\mu_c$ is always equal to the mass of the meson for the condensation. Therefore, we observe only a $\omega$ dependency of $\mu_c$ as shown in Fig.~\ref{fig2}. With $\omega$, the $\mu_c$ decreases, comes to a minimum, and then increases slowly at a very high magnitude of rotation. Note that with the boundary condition of $R$ = 5 fm, the causality condition allows us to have a maximum $\omega$ value of 0.2 fm$^{-1}$ ($\sim$ 0.0394 GeV). Moreover, the $\mu_c$ never approaches zero even though $\omega$ is too high. This implies that only rotation can not induce the macroscopic BEC phenomena, even though it enhances them by requiring low $\mu$ to achieve BEC.  The combined effect of rotation and chemical potential (say, effective chemical potential) along with the fixed density leads to the observed behavior of $T_c$. In Ref.~\cite{Siri:2024cjw}, the authors have studied the effect of rotation on the thermodynamic properties of a BEC. In a rotating gas, the critical temperature is found to be less than that of a non-rotating case. However, the critical temperature for the BEC transition increases with rotation as it is directly proportional to the angular velocity considered.

\begin{figure*}[ht!]
\begin{center}
\includegraphics[scale = 0.44]{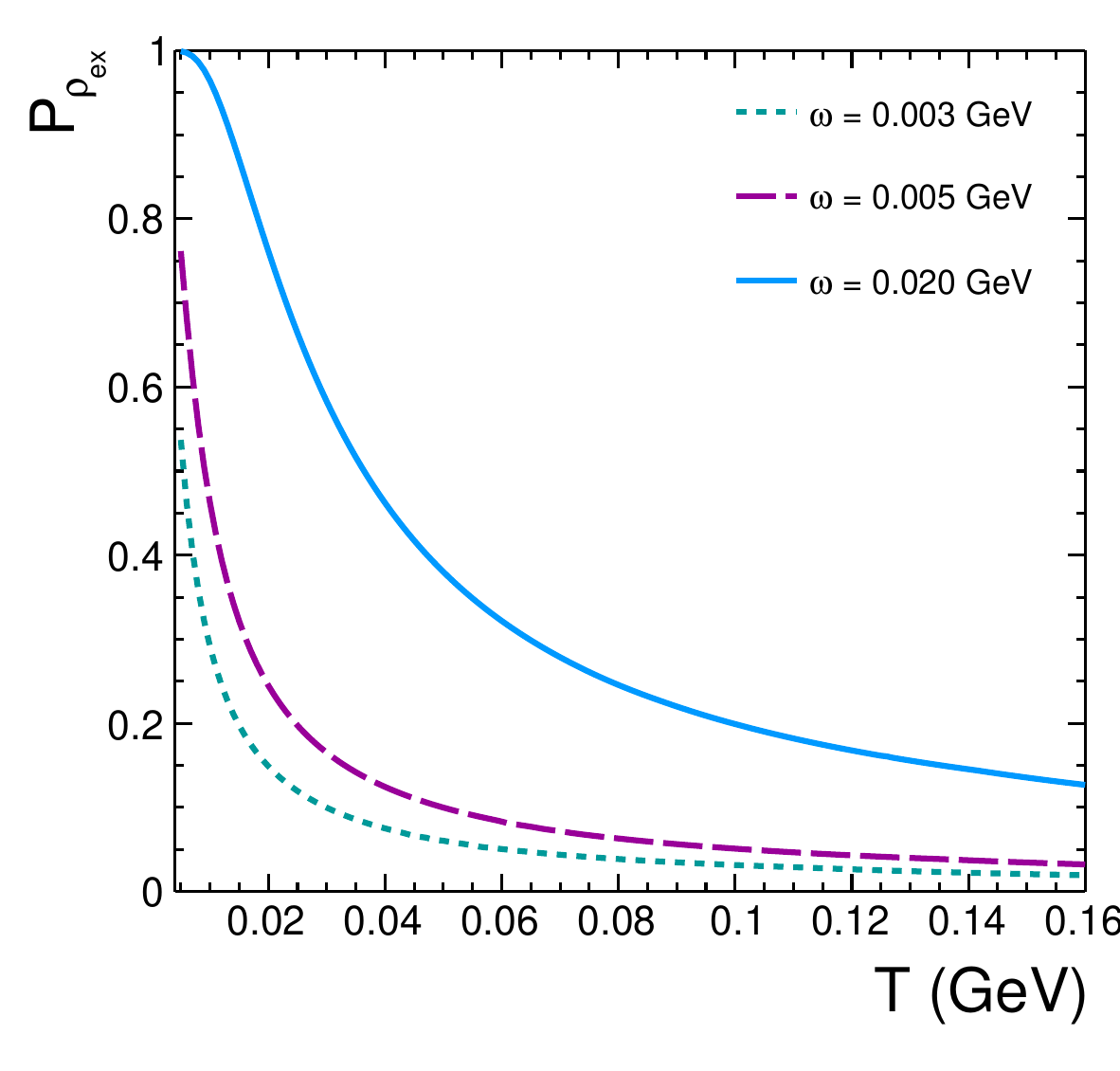}
\includegraphics[scale = 0.44]{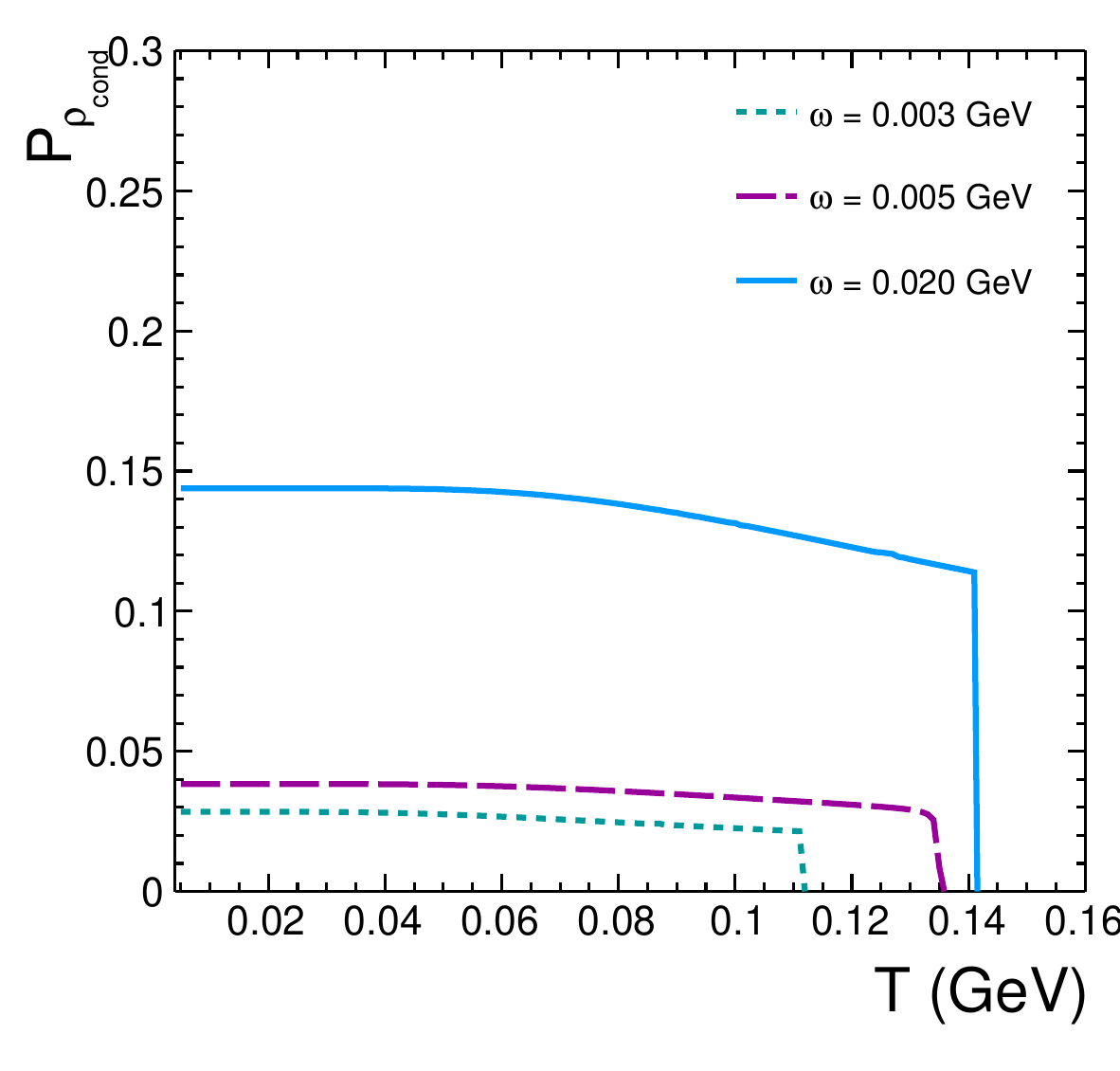}
\includegraphics[scale = 0.44]{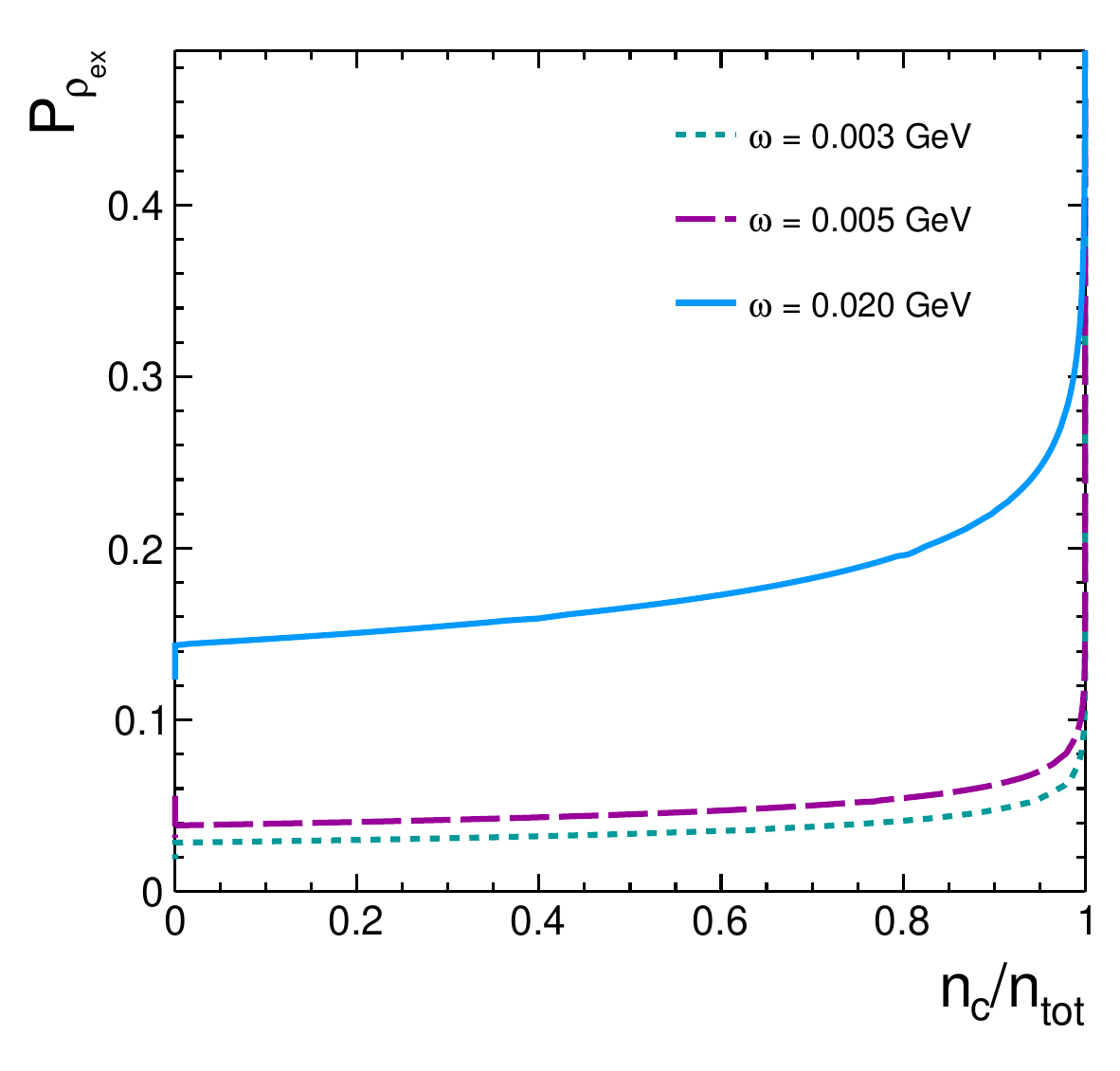}
\includegraphics[scale = 0.44]{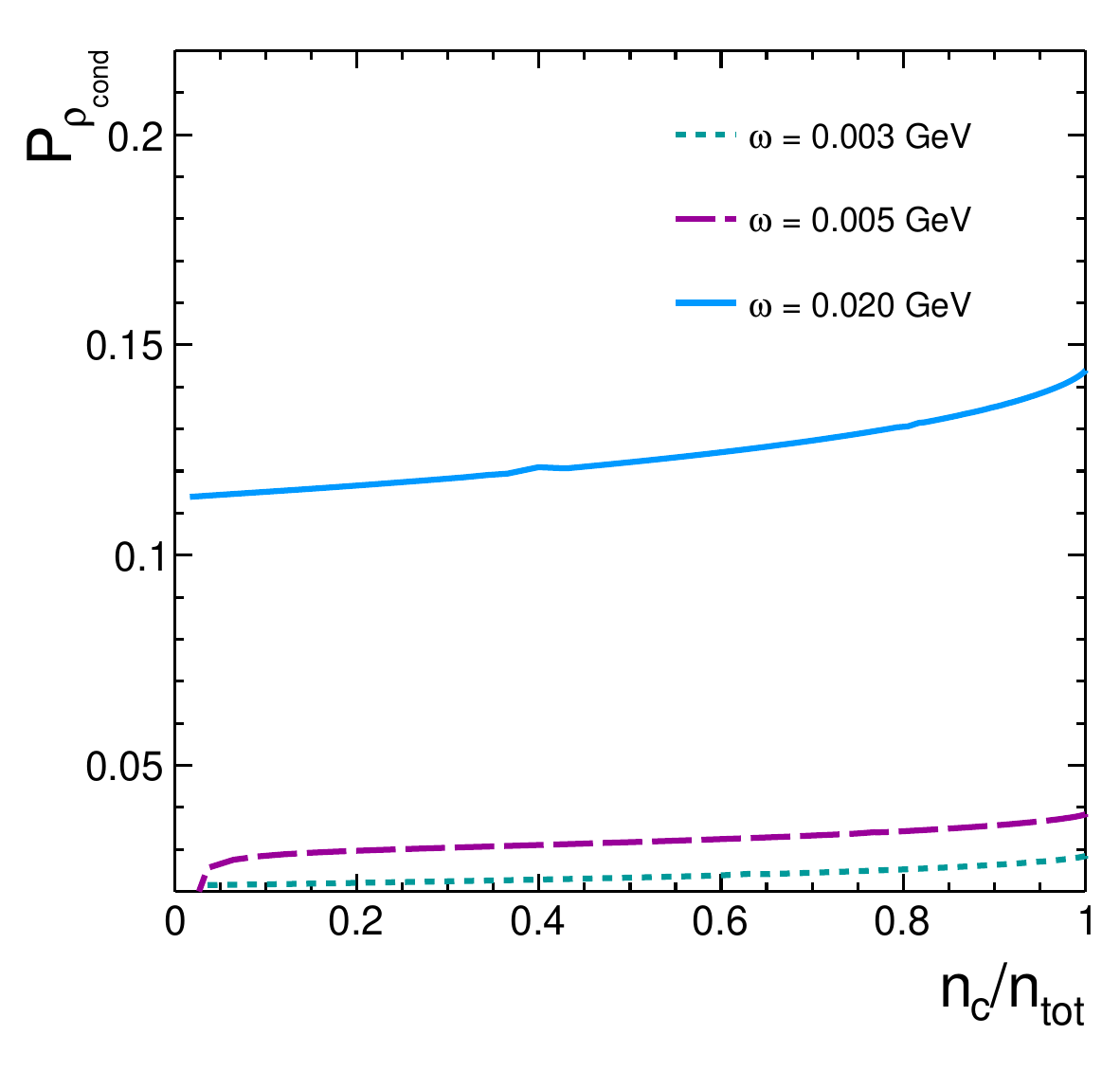}
\caption{(Colour Online) In the upper panels, the spin polarization of thermal (left) and condensate rho mesons (right) as a function of temperature is shown for different values of $\omega$. In the lower panel, the same is shown as a function of the condensate fraction.}
\label{fig3}
\end{center}
\end{figure*}

We examine the density of each spin state of the rho meson gas in both the condensate and the excited states. The spin alignment or the polarization of the particle can therefore be described by estimating the net spin density normalized by the total spin density as
\begin{equation}
    \label{eq_polarization}
    P_{\rho} = \frac{n_{\uparrow}-n_{\downarrow}}{n_{\uparrow}+n_{\downarrow}},
\end{equation}
where the $n_{\uparrow}$ and $n_{\downarrow}$ are the densities of up and down spin states of $\rho$ mesons, respectively. The polarization defined in Eq.~(\ref{eq_polarization}) quantifies the net spin alignment of the $\rho$ meson ensemble along the global rotation axis and represents the vector polarization, reflecting the imbalance between the populations of the spin-up and spin-down substrates. For a specific value of rotation, the sum of the spin density of each state at any temperature equals the total density of the $\rho$ gas considered. With rotation, there is an imbalance in the spin-up ($n_{\uparrow}$) and spin-down density ($n_{\downarrow}$), and these are separately fixed due to the constraint of fixed total density. However, these fixed populations are redistributed between the condensate and excited fractions as the temperature varies across the BEC transition. Consequently, the temperature dependence of the polarization arises solely from the evolving condensate and thermal fractions, while the total polarization remains determined exclusively by the rotation. Therefore, using Eq.~(\ref{eq_polarization}), we estimate the polarization separately for the excited ($P_{\rm ex}$) and condensate state ($P_{\rm cond}$) to study their behaviour under BEC. Since at $\omega$ = 0 GeV, all spin states have the same density, we investigate the polarization by considering three finite values of $\omega$: $\omega$ = 0.003, 0.005, and 0.02 GeV. Although experimental estimations from global hyperon polarization~\cite{STAR:2017ckg} show that $\omega$ varies in the range from $\sim$ 0.003 to 0.015 GeV, to investigate the effect of strong rotation on the dynamics of the polarization of the condensed and excited state, we also consider $\omega$ as high as 0.02 GeV. In the upper panel of Fig.~\ref{fig3}, the polarization results for excited and condensate particles are shown as a function of temperature. We observe a finite polarization for the excited density, which is maximum at low temperatures and decreases with increasing temperature. The polarization increases in magnitude with a higher rotation magnitude, being larger for the case of $\omega$ = 0.02 GeV within the considered temperature range. As $ T\to 0$, polarization for the excited particles is maximum, although the fraction of excited particles is also very less, as evident from Fig.~\ref{fig1}. At very high $\omega$, the density of excited particles is negligible; moreover, the density of down states is even much smaller, and whatever the excited density is, it is mostly due to spin up states. Therefore, at high $\omega$, almost fully polarized rho gas is observed at very low $T$. On the other hand, just at the onset of BEC, the condensate has a finite polarization which increases in magnitude with an increase in the value of $\omega$. As one approaches $T\rightarrow 0$, the condensate polarization increases slowly (with increasing condensate density as in Fig.~\ref{fig1}).

%In fig.~\ref{fig2} we have plotted the variation of BEC critical temperature as a function of rotation. The critical temperature $T_c$ for Bose-Einstein condensation exhibits a non-monotonic hump as a function of $\omega$ due to a competition between two effects. At low $\omega$, rotation weakens the effective confining potential, allowing the gas to expand and condense at a higher temperature, thus raising $T_c$. However, as $\omega$ increases further and approaches the transverse trapping frequency, the system enters a regime where the Coriolis force quantizes the particle motion into Landau levels. This drastically reduces the available energy states (density of states), making it far more difficult for a macroscopic occupation to form and thereby suppressing $T_c$ at high rotation speeds. The transition from the trap-softening regime to the Landau-level-dominated regime creates the characteristic peak in the $T_c$ versus $\omega$ curve.

We must keep in mind that $n_{\rm ex}$ and $n_{\rm cond}$ represent different populations with potentially different polarization states. While $n_{ex}$ will get polarized due to $\omega$ and $T$ effects, $n_{\rm cond}$ can have coherent polarization just from the macroscopic occupation of the ground state. The condensate itself is described by a macroscopic wavefunction. We define its polarization $P_{\text{cond}}$ as the spin alignment of this coherent ground state. In the lower panel of Fig.~\ref{fig3}, the polarization of the excited and condensate phases is obtained as a function of total condensate fraction. Our results show that the polarizations of both the condensate $n_{\text{cond}}$ and the excited states $n_{\text{ex}}$ increase gradually with the condensate fraction $n_{\text{cond}}/n_{\text{total}}$. This indicates that the emergence of Bose-Einstein condensation actively amplifies the global spin alignment beyond a simple population of polarized states. The condensate, once formed, acts as a source of a coherent mean field that breaks the spin-rotation symmetry. This field not only defines the condensate's own polarization but also exerts an aligning torque on the thermal cloud, resulting in the observed concurrent increase in $P_{\text{ex}}$. The condensate fraction $n_{\text{c}}/n_{\text{tot}}$ thus serves as a direct control parameter for the overall polarization strength of the system, unifying the behavior of both condensed and non-condensed components. It is noteworthy to mention that we have assumed a non-interacting $\rho$ gas for our study. The interaction among the particles may lead to a modification of the chemical potential, which can shift the value of $T_c$ from that of a non-interacting case. Also, in a highly dense interacting medium, there is a possibility of the reduction of the in-medium mass of the vector mesons, which may lead to a decrease in $\mu_c$ required for the BEC transition~\cite{Mallick:2014faa, Aharony:2007uu}. However, the mechanism due to rotation, such as the rotational enhancement of $T_{\rm c}$ and its effect on polarization, remains relatively the same. In this study, we present a foundational study using an ideal gas model to isolate and establish the novel interplay between rotation, BEC, and spin polarization. A detailed study involving an interacting medium could be done to understand the medium effect explicitly.

%In this work, we present the first unified study of Bose-Einstein condensation and spin polarization for $\rho$ mesons in a dense, rotating medium. %We demonstrate that global rotation not only enhances the condensation but also imprints a definitive polarization signature, which is further amplified by the coherent field of the condensate itself. 
The results obtained in this study can have direct relevance to the internal conditions of rotating neutron stars, which naturally combine high baryon density with rapid global rotation. Neutron stars are born with extremely high angular velocities, with initial rotation periods of the order of milliseconds, corresponding to $\omega \approx 10^{-2}$ GeV, and they retain significant rotational energy throughout much of their lifetime~\cite{Haensel:2007yy, Paschalidis:2016vmz}. We have demonstrated that rotation actively promotes the onset of Bose–Einstein condensation by effectively lowering the critical chemical potential ($\mu_c$) required for the transition. This behavior highlights the role of rotational dynamics as a catalyst for mesonic condensation in compact astrophysical objects. As predicted by various effective field theories, the in-medium mass of the $\rho$ meson, $m_\rho^*$, decreases with increasing baryon density in the presence of a strong magnetic field, bringing the $\rho$ meson energy spectrum closer to the chemical potential~\cite{Mallick:2014faa, Brauner:2016lkh, Shivam:2019cmw}. The high magnitude of rotation will again enhance the probability of achieving the condition for condensation. Moreover, the finite polarization of the condensate suggests the formation of a spin-aligned $\rho$ meson condensate within the neutron star core. The pressure exerted now differs parallel and perpendicular to the rotation (spin-polarization) axis, breaking spherical symmetry and inducing anisotropic deformation of the star. %breaking spherical symmetry and inducing anisotropic deformation of the star. % A spin-polarized condensate constitutes an anisotropic medium. The pressure exerted by this phase differs parallel and perpendicular to the rotation (spin-polarization) axis, breaking spherical symmetry and inducing anisotropic deformation of the star.

\section{Conclusion}

In this work, we present the first systematic study of the interplay between rotation, Bose-Einstein condensation, and spin polarization for $\rho$ mesons in a dense, hadronic system. Our results demonstrate that global rotation not only enhances $\rho$-meson condensation by lowering the critical chemical potential $\mu_c$ but also induces a strong spin alignment of the condensate. A key finding is that the onset of BEC actively amplifies this polarization, as the macroscopic wavefunction of the condensate acts as a coherent mean field that breaks spin-rotation symmetry and aligns both the condensed and thermal components.
Our results have direct implications for two frontiers of physics. In the context of heavy-ion collisions, the predicted large polarization anisotropy should be observable as a characteristic modulation in the angular distribution of decay daughters.  
When the system has high vorticity of the order of $\omega \sim $ 0.015 GeV~\cite{STAR:2017ckg}, our analysis shows that the critical chemical potential required for the onset of $\rho$ meson condensation is lowered to roughly two-fifths of the $\rho$ meson mass. Such a combined condition of moderately large chemical potential and vorticity is naturally achieved in heavy-ion collisions at low centre-of-mass energies. Using the established parameterizations relating chemical potential and center-of-mass energies~\cite{Cleymans:2005xv, Hatta:2015hca}, one finds that heavy-ion collisions in the range $\sqrt{s_{\rm NN}}\sim$ 3 to 10 GeV, accessible at RHIC, FAIR, and NICA can result in high chemical potential and vorticity, and may provide suitable conditions to study the effect of BEC phenomena in the $\rho$ meson polarization. For neutron stars, the presence of a spin-polarized $\rho$-meson condensate would not only soften the equation of state but could also generate an anisotropic pressure, potentially leading to observable macroscopic effects such as stellar deformation. %, potentially leaving imprints on the pre- and post-merger gravitational wave signals from binary neutron star mergers.
%the required chemical potential is reduced to almost two-fifths of the mass of the $\rho$ meson, which is achievable at heavy-ion collisions with low centre of mass energy, such as at RHIC, FAIR, and NICA energies. Therefore, considering the parametrized relations for chemical potential and center of mass energies~\cite{Cleymans:2005xv, Hatta:2015hca}, one can see that the low-energy heavy-ion collisions with $\sqrt{s_{\rm NN}}\sim$ 3 to 10 GeV can result in high chemical potential and vorticity and may provide suitable conditions to study the effect of BEC phenomena in the $\rho$ meson polarization.

\section*{Acknowledgement}

KKP  acknowledges the financial aid from UGC, Government of India. The authors gratefully acknowledge the DAE-DST, Govt. of India funding under the mega-science project -- “Indian participation in the ALICE experiment at CERN" bearing Project No. SR/MF/PS-02/2021-IITI (E-37123). DS acknowledges the support from the postdoctoral fellowship of DGAPA UNAM.

\end{document}